\documentclass[aps,prl,
superscriptaddress,reprint]{revtex4-1}
\usepackage{graphicx}
\usepackage{setspace}
\usepackage{amsmath}
\usepackage{amssymb}
\usepackage{color}
\usepackage{array}
\usepackage{subfigure}
\usepackage{hyperref}
\usepackage{float}
\usepackage{lipsum}
\usepackage{ulem}
\usepackage[all]{xy}
\newcommand{\RN}[1]{%
  \textup{\uppercase\expandafter{\romannumeral#1}}%
}

\begin{document}

\title{Representation of fermions in the Pati-Salam model}
\author{De-Sheng Li}, 
\affiliation{College of Electrical and Information Engineering, Hunan Institute of Engineering, Xiangtan 411100, P.R.China}
\affiliation{College of Physics, Mechanical and Electrical Engineering, Jishou University, Jishou 416000, P.R.China}
\author{Hong-Fei Zhang}
\affiliation{College of Big Data Statistics, Guizhou University of Finance and Economics,
Guiyang, 550025, P.R.China}

\begin{abstract}
In this paper, a representation of fermions in the Pati-Salam model is suggested. The semi-leptonic and beyond standard model flavor changing neutral currents of the Lagrangian in this representation of fermions are discussed. A pair of possible Cabibbo-Kobayashi-Maskawa and Pontecorvo-Maki-Nakagawa-Sakata matrices are defined. An effective Lagrangian for this model is given.

\end{abstract}
\maketitle


\section{Introduction}
\label{intro}
The Pati-Salam model \cite{Pati:1974yy} is a grand unified theory (GUT) \cite{Pati:1974yy, Weinberg:1967tq, Georgi:1974sy, Langacker:1980js, Antoniadis:1987dx, Antoniadis:1989zy} and has the gauge group structure of $SU(4)_{L}\times SU(4)_{R} \times SU(4^{\prime})$, where $SU(4)_{L}\times SU(4)_{R}$ is the chiral flavor gauge group, and $SU(4^{\prime})$ is the color gauge group.
The gauge group structure of the Pati-Salam model is beneficial in several aspects
\begin{itemize}
\item The minimal simple group $SU(5)$ GUT \cite{Georgi:1974sy} encounters the issue of proton decay, and the modifications used to address the proton decay problem in $SU(5)$ GUT always encounter issues of naturalness.
\item If we use a semi-simple group as the GUT gauge group instead of a simple group, the standard model (SM) particles phenomena could be unified with the Pati-Salam gauge group $SU(4)_{L}\times SU(4)_{R} \times SU(4^{\prime})$, where $SU(4)_{L}\times SU(4)_{R}$ is the chiral flavor gauge group, and $SU(4^{\prime})$ is the color gauge group. While the gauge group $SU(2)_{L}\times SU(2)_{R}\times SU(4^{\prime})$ model could be used to reproduce the neutral current (NC) and charge current (CC) weak interaction phenomena, the six flavor fermions and flavor mixing phenomena are difficult to reproduce.

\item ``Lepton number as the fourth color" \cite{Pati:1974yy} is a clean and straightforward assumption when visualizing the fermions from a unified viewpoint. 
\item The fundamental representations of $SU(4)$ are $\mathbf{4}$, $\mathbf{6}$ and $\mathbf{\bar{4}}$. In a GUT, the fermions always fill in the fundamental representation of a gauge group. We know that fermions have six flavors and four colors, and each fermion has corresponding antifermion. Thus, fermions (antifermions) can be filled in the Pati-Salam gauge group fundamental representation $\mathbf{4}\times \mathbf{6}$ ($\mathbf{\bar{4}}\times \mathbf{6}$). 
\item Dirac matrices are $4\times 4$ matrices. If we do not add (or reduce) the degrees of freedom by hand, the fermions should fill in the $4\times 4$ matrix. 
\item The flavor mixing matrices, i.e., Cabibbo-Kobayashi-Maskawa (CKM) and Pontecorvo-Maki-Nakagawa-Sakata (PMNS) matrices, could arise naturally from $SU(4)_{L}\times SU(4)_{R} \times SU(4^{\prime})$ Pati-Salam model.
\item Pati-Salam model \cite{Pati:1974yy} as the flat spacetime limits of Pati-Salam model in curved spacetime, can be derived from self-parallel transportation principle of square root Lorentz manifold \cite{Li:2018fxl}, which is a pure geometry model. An explicit formulation of sheaf quantization \cite{Isham:1999kb,Doering:2007syn,kashiwara2013sheaves,Nakayama:2014tga,flori2018second,Kuwagaki:2022rpj} on square root Lorentz manifold is given and the relation between sheaf quantization and path integral quantization is shown \cite{Li:2022vaw}, the canonical quantization of Yang-Mills theory in curved spacetime which inspired by sheaf quantization can be seen \cite{Li:2023izg} also. The abstract category structure of sheaf quantization of square root Lorentz manifold almostly like Lagrangian submanifold on symplectic geometry \cite{Fukaya:2011ad,Esen:2021enx}. 
\end{itemize}

Gauge group structure $SU(4)_{L}\times SU(4)_{R} \times SU(4^{\prime})$ of the Pati-Salam model is the starting point of this paper. In an existing paper \cite{Pati:1974yy}, the chiral flavor group $SU(4)_{L}\times SU(4)_{R}$  degenerates into the chiral group $SU(2)_{L}\times SU(2)_{R}$ and reproduces the NC and CC weak interactions transported by $Z$ and $W^{\pm}$ weak gauge bosons, respectively. 
The left-right symmetry of the Pati-Salam model predicts the existence of right handed neutrinos. The $SU(4^{\prime})$ color group from the conjecture ``lepton number as the fourth color'' contains $SU(3^{\prime})$ quantum chromodynamics (QCDs) and exotic semi-leptonic processes transported by $X$ bosons. The semi-leptonic processes preserve $B-L$ symmetry and violate baryon lepton number conservation. Topics such as $B-L$ symmetry \cite{Davidson:1978pm,Mohapatra:1980qe}, baryogenesis \cite{Weinberg:1979sa,Vergados:1985pq,Krauss:1999ng,Riotto:1999yt,Dimopoulos:1978kv, Marshak:1979fm,Weinberg:1980bf,Perez:2014qfa,deGouvea:2014lva,Dev:2020qet}, leptogenesis \cite{Weinberg:1980bf,Pilaftsis:2003gt,Fong:2013wr,Cai:2017mow,Chun:2017spz,Zhang:2020lir,Xing:2020ald}
, left-right symmetry \cite{Chang:1984uy}, and right handed neutrinos \cite{Drewes:2013gca} have been
important topics in theoretical and experimental high energy physics for decades. Recent 
literature has discussed the flavor violation \cite{Ilakovac:1994kj,Feruglio:2008ht,Gavela:2009cd,Alonso:2011jd,Han:2012vk,deGouvea:2013zba,Xing:2015fdg,Xing:2019vks,Feruglio:2019ktm,Barducci:2020ncz,Deppisch:2020oyx,Husek:2020fru,RochaMoran:2020cqp,Fuentes-Martin:2020hvc,Ellis:2020jfc}, neutral gauge boson \cite{Leike:1998wr,Altarelli:1989ff,Huong:2019rxc}, lepton quark collider \cite{Buchmuller:1986zs}, lepton flavor universality\cite{Bryman:2021teu}, gravitational wave imprints \cite{Graf:2021xku}, muon $g-2$ anomaly \cite{Perez:2021ddi}, and muon collider \cite{Asadi:2021gah} which relate to the Pati-Salam model and other models.

The original fermions representation in the Pati-Salam model \cite{Pati:1974yy} includes only two families of quarks and leptons. In this paper, however, we suggest a representation of fermions in the Pati-Salam model comprising all three families of quark and lepton states as the eigenstates of Lagrangians. We discuss the fermion-antifermion-boson vertices new physics of semi-leptonic processes transpoted by $X$ bosons and beyond standard model flavor changing neutral currents (FCNCs) processes transported by neutral bosons $Y$, based on the novel representation of fermions. We also present a possible construction of the CKM and PMNS matrices based on this representation of fermions. Finally, we illustrate an effective total Lagrangian density for this model.

\section{Representation of fermions }
The well-established Pati-Salam model \cite{Pati:1974yy} has the following gauge group
\begin{eqnarray}\label{gaugegroup}
G= SU(4)_{L}\times SU(4)_{R} \times SU(4^{\prime})
\end{eqnarray}
where $SU(4)_{L}$ and $SU(4)_{R}$ are the chiral flavor gauge groups, $SU(4^{\prime})$ is the color group.

Fermions have six flavors of quarks and leptons. If we gauge the flavor symmetry according to $SU(6)$ group, the fermions should fill in a $4\times 6$ matrix. The $SU(6)$ flavor symmetry will engage with nine gauge bosons at least that transport flavor gauge interactions. To date, the experimental data only showed us three flavor gauge interaction bosons, which are $W^{+}, W^{-}$ and $Z$. The problem relates to how to reduce the nine flavor gauge bosons naturally to three, reveal the Standard Model interaction vertexes and reproduce flavor mixing phenomena. Furthermore, it will be hard to reproduce the Gell-Mann-Nishijima formula and flavor mixing phenomena, and the $SU(6)\times SU(4^{\prime})$ gauge group is not minimal for GUT.

This $SU(4)_{L}\times SU(4)_{R}$ flavor gauge group symmetry restricts the representation matrix of fermions to $4\times 4$ matrix. For this $4\times 4$ fermion matrix, it needs to be established whether the flavor degrees of freedom will take on the shape of a column or row? A minimal coupling Lagrangian is constructed as follow for the color and flavor interaction to answer this question 
\begin{eqnarray}
\label{Lcolor}
\mathcal{L}&=& \mathbf{Tr}\left[i\bar{\Psi}\gamma^{\mu}\partial_{\mu}\Psi+f \bar{\Psi} \gamma^{\mu}V_{\mu}\Psi-g\bar{\Psi}\gamma^{\mu} \Psi W_{\mu}\right],
\end{eqnarray}
where $f,g\in \mathbb{R}$ are coupling constants. $V_{\mu}$ and $W_{\mu}$ are $4\times 4$ Hermitian matrices and can be decomposed as follows
\begin{eqnarray}
V_{\mu}=\sum_{a=1}^{15}V_{\mu}^{a}T^{a},&\ \ \ & W_{\mu}=\sum_{a=1}^{15}W_{\mu}^{a}T^{a},
\end{eqnarray}
where $T^{a} (a=1,2,\cdots,15)$ are generators of $SU(4)$ and an example can be found in Appendix A, $V_{\mu}^{a}$ and $W_{\mu}^{a}$ are gauge bosons.
The first term in Lagrangian (\ref{Lcolor}) is a kinematic term. The flavor interaction can be chiral decomposed but the color interaction cannot. We observe that the second term in Lagrangian (\ref{Lcolor}) is difficult to decompose due to chiral symmetry, but the third term  can be decomposed (the proof is presented in Appendix {\ref{AppendixB}}) as follow
\begin{eqnarray*}
\label{color}
\mathcal{L}=\mathbf{Tr}\left[i\bar{\Psi}\gamma^{\mu}\partial_{\mu}\Psi+\sum_{a=1}^{15} \left(f\bar{\Psi} \gamma^{\mu}V_{\mu}^{a}{T}^{a}\Psi-g\bar{\Psi}_{L}\gamma^{\mu} \Psi_{L} W_{\mu}^{a}{T}^{a}
\right.\right.\\  \left.\left.
-g\bar{\Psi}_{R}\gamma^{\mu} \Psi_{R} W_{\mu}^{a}{T}^{a}\right)\right],
\end{eqnarray*}
where the chiral fermions are defined
\begin{eqnarray}
\Psi_{L}=\frac{1-\gamma^{5}}{2} \Psi, &\quad&
\Psi_{R}=\frac{1+\gamma^{5}}{2} \Psi,\\
\bar{\Psi}_{L}=\Psi^{\dagger} \frac{1-\gamma^{5}}{2}\gamma^{0} , &\quad&
\bar{\Psi}_{R}=\Psi^{\dagger} \frac{1+\gamma^{5}}{2} \gamma^{0}.
\end{eqnarray}
Accordingly, the second term in Lagrangian (\ref{Lcolor}) describes the $SU(4^{\prime})$ color gauge interaction, and the third term in Lagrangian (\ref{Lcolor}) describes the $SU(4)_{L}\times SU(4)_{R}$ chiral flavor gauge interaction. We then derive that the column of the $4\times 4$ fermion matrix corresponding to color and the row corresponding to flavor.

Such as ``lepton number as the fourth color'', it was then easy to fill four colors of fermions, i.e., $R,G,B$ and $L$, into the four rows of fermion matrix. The next approach was to derive how to fill the six flavor fermions into the four columns of the fermion matrix? Reminding the six flavor fermions were divided into three families, and each family included two flavor fermions. The action in the path integral formulation of quantum field theory is a phase
\begin{eqnarray}
S=\int d^{4}x \mathcal{L},
\end{eqnarray}
each phase term should with $0$-dimension and $0$-charge, and the fermion matrix should result in the model being anomaly free. Consider that the fermions in quantum field theory are the operator valued field, and the quantum states are the eigenstates of operator valued field. In quantum mechanics, one operator can correspond to several eigenstates. Then, we suggest a representation of fermions
\begin{eqnarray}
\Psi=\left(\begin{array}{cccc}
\sqrt{2}u_{R}& \sqrt{2}c_{R} & \sqrt{2}t_{R} &d^{\prime}_{R}\\
\sqrt{2}u_{G}& \sqrt{2}c_{G} & \sqrt{2}t_{G} &d^{\prime}_{G}\\
\sqrt{2} u_{B}& \sqrt{2}c_{B} & \sqrt{2}t_{B} &d^{\prime}_{B}\\
e & \mu &\tau & \nu^{\prime}
\end{array}\right),
\end{eqnarray}
where $C=R,G,B=1,2,3$ are color indices; $u,c$ and $t$ are operator valued fields of three flavor quarks; $e,\mu$ and $\tau$ are operator valued fields of the electron, mu and tau. Furthermore, $\nu^{\prime}$ and  $d^{\prime}_{C}$ are operator valued fields of neutrinos and $d$ family quarks. Additionally, the $|\nu_{e}\rangle, |\nu_{\mu}\rangle, |\nu_{\tau}\rangle$ neutrino states and $|d_{C}\rangle, |s_{C}\rangle, |t_{C}\rangle$ quark states are eigenstates related to flavor interaction Lagrangian terms containing $\nu^{\prime}$ and $d^{\prime}_{C}$, respectively.

\section{Gauge bosons in the minimal coulping Lagrangian}
The possibility of chiral decomposition infers that $W_{\mu}^{a}$ are gauge bosons transporting flavor gauge interactions and $V_{\mu}^{a}$ transporting color gauge interactions. We will discuss the decomposition of the Lagrangian of the flavor and color interactions in detail using the minimal coulping model (\ref{Lcolor}).
\begin{figure}
\begin{center}
\includegraphics[width=86 mm]{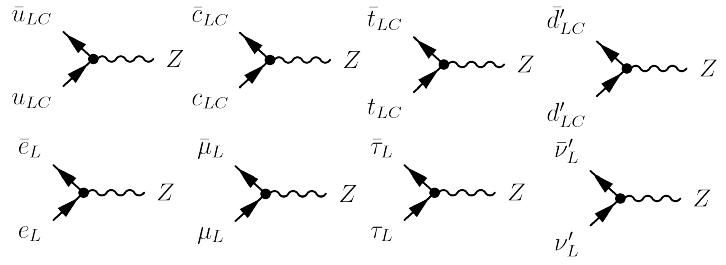}
\caption{\label{picZ}In this suggested representation of fermions of Pati-Salam model, the Lagrangian (\ref{Z}) gives us the fermion-antifermion-boson vertexes of weak $Z$ boson interaction with left handed fermions. For right handed fermions, the $L$ symbol should be alternated by $R$. }
\end{center}
\end{figure}

\subsection{Chiral flavor $SU(4)_{L}\times SU(4)_{R}$ processes}
The gauge boson bears the exchange of quantum numbers charge. For two different fermion-antifermion-boson vertexes, when the exchange of the charge is the same, the quantum numbers of two gauge bosons in two fermion-antifermion-boson vertexes are the same, except the possibility of masses difference (thanks the comments from anonymous referees point out that even through the quantum number of the particles are the same, the masses of the particles might not the same). The Z boson is a charge free gauge boson and transports weak NC in the SM, Z boson should on the diagonal of matrix $W_{\mu}$, i.e.,
\begin{eqnarray}
Z_{\mu}=W^{3}_{\mu}=W^{8}_{\mu}=W^{15}_{\mu},
\end{eqnarray}
then the Lagrangian
\begin{eqnarray}
-g\mathbf{Tr}\left[\bar{\Psi}_{L}\gamma^{\mu} \Psi_{L} \sum_{a=3,8,15} W_{\mu}^{a}{T}^{a}+\{ L\rightarrow R\}\right]
\end{eqnarray}
can be decomposed as follows (see Fig.~\ref{picZ})
\begin{eqnarray} \label{Z}\nonumber
&&-g\mathbf{Tr}\left[\bar{\Psi}_{L}\gamma^{\mu} \Psi_{L} \sum_{a=3,8,15} W_{\mu}^{a}{T}^{a}+\{ L\rightarrow R\}\right]\\ \nonumber
=&&-g \mathbf{Tr}\left[ \sum_{C=R,G,B} \left(  \zeta_{1}\bar{u}_{LC}\gamma^{\mu}u_{LC} Z_{\mu}+\zeta_{2}\bar{c}_{LC}\gamma^{\mu}c_{LC} Z_{\mu}\right.\right. \\  \nonumber
&&\left. +\zeta_{3}\bar{t}_{LC}\gamma^{\mu}t_{LC} Z_{\mu}\right)   +   \frac{1}{2}\left(\zeta_{1}\bar{e}_{L}\gamma^{\mu}e_{L} Z_{\mu}  +\zeta_{2}\bar{\mu}_{L}\gamma^{\mu}\mu_{L} Z_{\mu} \right. \\   \nonumber
&&\left.+\zeta_{3}\bar{\tau}_{L}\gamma^{\mu}\tau_{L} Z_{\mu} +\zeta_{4}\bar{\nu}^{\prime}_{L}\gamma^{\mu}\nu^{\prime}_{L} Z_{\mu}\right)\\   
&&\left.+\zeta_{4} \sum_{C=R,G,B}\bar{d}^{\prime}_{LC}\gamma^{\mu}d^{\prime}_{LC} Z_{\mu}+\{L \rightarrow R\}\right],
\end{eqnarray}
where 
\begin{eqnarray*}
\zeta_{1}=1+\frac{\sqrt{3}}{3}+\frac{\sqrt{6}}{6},&\ \ \ &
\zeta_{2}=-1+\frac{\sqrt{3}}{3}+\frac{\sqrt{6}}{6},\\
\zeta_{3}=-\frac{2\sqrt{3}}{3}+\frac{\sqrt{6}}{6},&\ \ \ &
\zeta_{4}=-\frac{\sqrt{6}}{2}.
\end{eqnarray*}

\begin{figure}
\begin{center}
\includegraphics[width=63 mm]{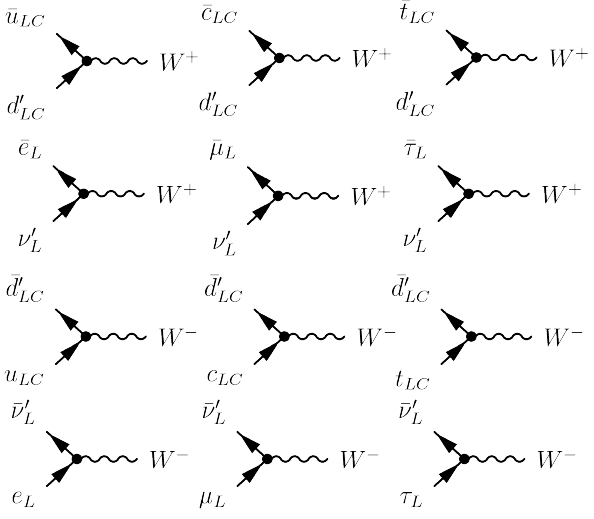}
\caption{\label{picW}The fermion-antifermion-boson vertexes of $W$ boson derived by Lagrangian (\ref{W}), where all three external legs of vertexes in this figure are momentum in.}
\end{center}
\end{figure}

According to the fermion matrix and Lagrangian charge free assumption, it is easy to find that $W^{\pm }_{\mu}$ in this model are
\begin{eqnarray*}
W^{\pm}_{\mu}=W^{9}_{\mu} \pm i W^{10}_{\mu}=W^{11}_{\mu} \pm i W^{12}_{\mu}=W^{13}_{\mu} \pm i W^{14}_{\mu}.
\end{eqnarray*}
 Furthermore, $W^{\pm}$ transports the CC in the weak interaction.
Then, the Lagrangian, i.e.,
\begin{eqnarray}\nonumber
-g \mathbf{Tr}\left[\bar{\Psi}_{L} \gamma^{\mu} \Psi_{L} \sum_{a=9}^{14}W_{\mu}^{a}{T}^{a}+\{ L\rightarrow R\}\right]
\end{eqnarray}
can be decomposed as follows (see Fig.~\ref{picW})
\begin{eqnarray}\label{W} \nonumber
&&-g \mathbf{Tr}\left[\bar{\Psi}_{L} \gamma^{\mu} \Psi_{L}  \sum_{a=9}^{14}W_{\mu}^{a}{T}^{a}+\{ L\rightarrow R\}\right]\\   \nonumber
&=&\frac{-g}{2} \mathbf{Tr}\left[\sqrt{2}\sum_{C=R,G,B} \left( \bar{u}_{LC} \gamma^{\mu} d^{\prime}_{LC} W^{+}_{\mu}+ \bar{c}_{LC} \gamma^{\mu} d^{\prime}_{LC} W^{+}_{\mu}\right. \right.\\ \nonumber
&& + \bar{t}_{LC} \gamma^{\mu} d^{\prime}_{LC} W^{+}_{\mu}+\bar{d}^{\prime}_{LC}\gamma^{\mu}u_{LC}W^{-}+\bar{d}^{\prime}_{LC}\gamma^{\mu}c_{LC}W^{-}\\  \nonumber
&&\left.+\bar{d}^{\prime}_{LC}\gamma^{\mu}t_{LC}W^{-} \right)+\bar{e}_{L} \gamma^{\mu} \nu^{\prime}_{L} W^{+}_{\mu} +\bar{\mu}_{L} \gamma^{\mu} \nu^{\prime}_{L} W^{+}_{\mu}\\  \nonumber
&&   +\bar{\tau}_{L} \gamma^{\mu} \nu^{\prime}_{L} W^{+}_{\mu}+\bar{\nu}^{\prime}_{L} \gamma^{\mu} e_{L} W^{-}_{\mu}+\bar{\nu}^{\prime}_{L} \gamma^{\mu} \mu_{L} W^{-}_{\mu}\\  
  &&\left.+\bar{\nu}^{\prime}_{L} \gamma^{\mu} \tau_{L} W^{-}_{\mu}+\{ L\rightarrow R\} \right].
\end{eqnarray}
The electric charge of $W^{+}$ and $W^{-}$ are $1$ and -1, respectively.

There are new physics chiral flavor processes described by the Lagrangian
\begin{eqnarray}\label{Y}  \nonumber
&&-g \mathbf{Tr}\left[\bar{\Psi}_{L} \gamma^{\mu} \Psi_{L}  \sum_{a=1,2,4,5,6,7}W_{\mu}^{a}{T}^{a}+\{ L\rightarrow R\}\right]\\  \nonumber
&=&{-g}\mathbf{Tr}\left[ \sum_{C=R,G,B} \bar{u}_{LC}\gamma^{\mu} (c_{LC}Y^{1}_{* \mu}+t_{LC}Y^{2}_{* \mu}) \right.\\  \nonumber
&&+\frac{1}{2}\bar{e}_{L}\gamma^{\mu}(\mu_{L}Y^{1}_{*\mu}+\tau_{L}Y^{2}_{*\mu}) +\frac{1}{2}\bar{\mu}_{L}\gamma^{\mu}(e_{L}Y^{1}_{\mu}+\tau_{L}Y^{1}_{*\mu}) \\  \nonumber
&& +\sum_{C=R,G,B} \bar{c}_{LC}\gamma^{\mu} (u_{LC}Y^{1}_{\mu}+t_{LC}Y^{1}_{*\mu})\\  \nonumber
&& +\sum_{C=R,G,B} \bar{t}_{LC}\gamma^{\mu} (u_{LC}Y^{2}_{\mu}+c_{LC}Y^{1}_{\mu})\\
&&\left.+\frac{1}{2}\bar{\tau}_{L}\gamma^{\mu}(e_{L}Y^{2}_{\mu}+\mu_{L}Y^{1}_{\mu})+\{L \rightarrow R\}\right].
\end{eqnarray}

\begin{figure}
\begin{center}
\includegraphics[width=86 mm]{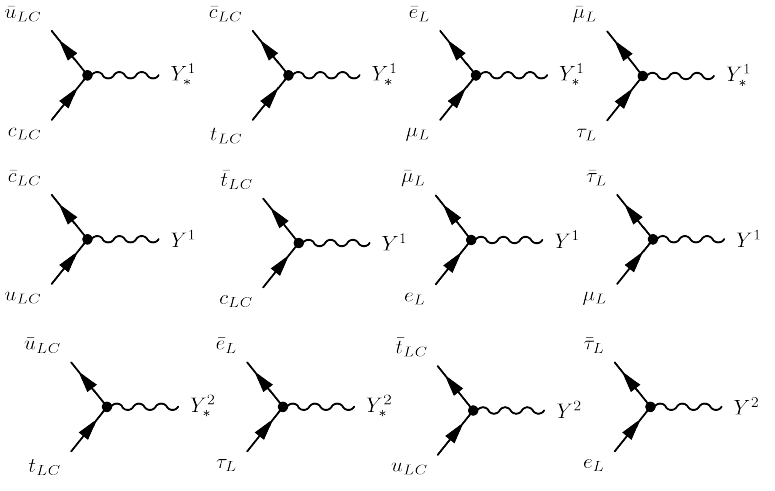}
\caption{\label{picY}The fermion-antifermion-boson vertexes of $Y$ are derived by Lagragian (\ref{Y}), where all three external legs of the vertexes are momentum in.}
\end{center}
\end{figure}

For example, the predicted beyond SM FCNCs \cite{vanVeghel:2020pmk, Martynov:2020cjd,Kumbhakar2020okw}
\begin{eqnarray}
Y^{1}_{*} \rightarrow u^{C}+\bar{c}^{C},\\
Y^{1}_{*} \rightarrow e^{+}+\mu^{-},
\end{eqnarray}
not yet being observed and the mass generating mechanism of gauge bosons $Y^{1}$, $Y^{2}$, $Y^{1}_{*}$ and $Y^{2}_{*}$ is interesting. The electric charges of gauge bosons $Y^{1}$, $Y^{2}$, $Y^{1}_{*}$ and $Y^{2}_{*}$ are 0. The fermion-antifermion-boson vertexes about $Y$ are show in Fig.~\ref{picY}. Two examples of beyond SM tree level FCNCs are in Fig.~\ref{pic3}.

The $W_{\mu}$ matrix is
\begin{eqnarray}
W_{\mu}=\frac{1}{2}\left(\begin{array}{cccc}
\zeta_{1} Z_{\mu}& Y_{\mu}^{1}&Y_{\mu}^{2}& W_{\mu}^{-}\\
Y_{*\mu}^{1}& \zeta_{2} Z_{\mu}&Y_{\mu}^{1}&W_{\mu}^{-}\\
Y_{*\mu}^{2}& Y_{*\mu}^{1}&\zeta_{3}Z_{\mu}&W_{\mu}^{-}\\
W_{\mu}^{+}& W_{\mu}^{+}&W_{\mu}^{+}& \zeta_{4}Z_{\mu}
\end{array}\right).
\end{eqnarray}
The corresponding electric charge matrix of $W_{\mu}$ is
\begin{eqnarray}
Q_{W}=\left(\begin{array}{cccc}
0& 0&0& -1\\
0& 0&0& -1\\
0& 0&0& -1\\
1& 1&1&0
\end{array}\right).
\end{eqnarray}

\subsection{Color $SU(4^{\prime})$ processes}
We selected $V^{15}_{\mu} $ as the the photon.
Then the vertexes of photon from Lagrangian \eqref{Lcolor} are written as
 \begin{eqnarray}\label{gamma}\nonumber
&&f \mathbf{Tr}\left[\bar{\Psi} \gamma^{\mu} V_{\mu}^{15}{T}^{15} \Psi\right]\\   \nonumber
  &=&f\frac{\sqrt{6}}{4}\mathbf{Tr}\left[ \frac{2}{3}\sum_{C=R,G,B}(\bar{u}_{C}\gamma^{\mu}V_{\mu}^{15}u_{C} +\bar{c}_{C}\gamma^{\mu}V_{\mu}^{15}c_{C} \right.
\\  \nonumber
&&+\bar{t}_{C}\gamma^{\mu}V_{\mu}^{15} t_{C} )-(\bar{e}\gamma^{\mu} V_{\mu}^{15} e   +\bar{\mu}\gamma^{\mu}V_{\mu}^{15} \mu +\bar{\tau}\gamma^{\mu}V_{\mu}^{15} \tau )\\
&&\left.+\frac{1}{3}\sum_{C=R,G,B}\bar{d}^{\prime}_{C}\gamma^{\mu}V_{\mu}^{15} d^{\prime}_{C} -\bar{\nu}^{\prime}\gamma^{\mu}V_{\mu}^{15} \nu^{\prime} \right].
\end{eqnarray}
Except the neutrinos, the electric charge number preseding each flavor fermion Lagrangian term is correct. As an example, the $\frac{2}{3}$ preceding the Lagragian term $\bar{u}_{C}\gamma^{\mu}V_{\mu}^{15}u_{C}$ is the electric charge number of quark $u$.
The experiments show that the neutrino is charge free, such that the neutrino should satisfy the formulas
\begin{eqnarray}\label{neutrino-r}
\nu^{\prime}=e^{i\theta^{\prime}}, \theta^{\prime\dagger}=\theta^{\prime}.
\end{eqnarray}
Under the restriction (\ref{neutrino-r}), the Lagrangian of neutrinos and photon interaction vertexes degenerates into
\begin{eqnarray}
-f\frac{\sqrt{6}}{4}\mathbf{Tr}\left[\bar{\nu}^{\prime}\gamma^{\mu}V_{\mu}^{15} \nu^{\prime} \right]=-f\frac{\sqrt{6}}{4}\mathbf{Tr}\left[\gamma^{0}\gamma^{\mu} V_{\mu}^{15}\right].
\end{eqnarray}

\begin{figure}
\begin{center}
\includegraphics[width=86 mm]{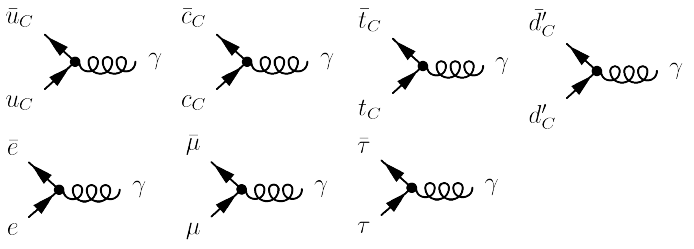}
\caption{\label{picgamma} The fermion-antifermion-boson verterxes of photon derived by Lagrangian (\ref{gamma}).}
\end{center}
\end{figure}

Then the fermion-antifermion-boson vertexes about photon $\gamma$ on this minimal coupling model are show in Fig.~\ref{picgamma}.

The gauge bosons $V^{1}_{\mu},V^{2}_{\mu},\cdots V^{8}_{\mu}$ are gluons and transport color $SU(3^{\prime})$ strong interaction and reveal QCD.

\begin{figure}
\begin{center}
\includegraphics[width=86 mm]{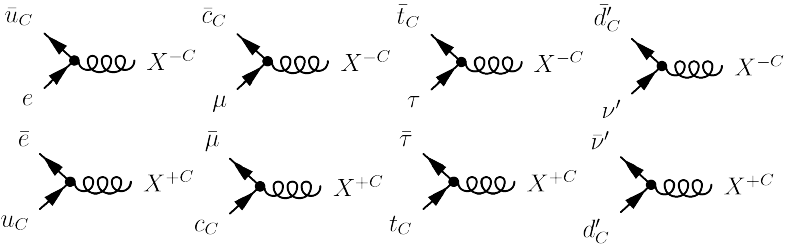}
\caption{\label{picX} The fermion-antifermion-boson veterxes derived by Lagrangian (\ref{Xboson}), where all three external legs of the vertexes are momentum in.}
\end{center}
\end{figure}

There are exotic semi-leptonic processes \cite{Fuentes-Martin:2020hvc} transported by $X^{\pm C}_{\mu}$ particles and the related Lagrangian is
\begin{eqnarray}    \label{Xboson} \nonumber
&&f \mathbf{Tr}\left[\bar{\Psi} \gamma^{\mu}\sum_{a=9}^{14}V_{\mu}^{a}{T}^{a}\Psi \right]\\   \nonumber
&=&f \frac{\sqrt{2}}{2} \sum_{C=R,G,B}  \mathbf{Tr}\left[\sqrt{2} \left( \bar{u}_{C} \gamma^{\mu} X^{-C}_{\mu}  e + \bar{c}_{C} \gamma^{\mu} X^{-C}_{\mu} \mu  \right.\right. \\  \nonumber
&&\left.+ \bar{t}_{C} \gamma^{\mu} X^{-C}_{\mu} \tau \right) +\bar{d}^{\prime}_{C} \gamma^{\mu} X^{-C}_{\mu} \nu^{\prime}   +\bar{\nu}^{\prime}\gamma^{\mu}X^{+C}d^{\prime}_{C}\\   
&&\left. +\sqrt{2}(\bar{e}\gamma^{\mu}X^{+C}u_{C}+\bar{\mu}\gamma^{\mu}X^{+C}c_{C}+\bar{\tau}\gamma^{\mu}X^{+C}t_{C}) \right],\ \ \ \ \ 
\end{eqnarray}
where 
\begin{eqnarray}
X^{\pm C}_{\mu}=V^{8+C}_{\mu}\pm iV^{9+C}_{\mu}.
\end{eqnarray}
The related fermion-antifermion-boson vertexes about $X$ bosons are show in Fig.~\ref{picX}.

The Lagrangian charge free restriction derives that the charge of $X^{-C}$ and $X^{+C}$ particles are $-\frac{1}{3}$ and $\frac{1}{3}$, respectively.
The $V_{\mu}$ matrix is
\begin{eqnarray}
V_{\mu}=\left(\begin{array}{cccc}
G^{RR}_{\mu}+V^{15}_{\mu}& G^{RG}_{\mu} & G^{RB}_{\mu} & X^{-R}_{\mu}\\
G^{GR}_{\mu}& G^{GG}_{\mu} +V^{15}_{\mu}& G^{GB}_{\mu} & X^{-G}_{\mu}\\
G^{BR}_{\mu}& G^{BG}_{\mu} & G^{BB}_{\mu} +V^{15}_{\mu}& X^{-B}_{\mu}\\
X^{+R}_{\mu}& X^{+G}_{\mu} & X^{+B}_{\mu}& -3V^{15}_{\mu}
\end{array}\right),\ \ \ \ \ 
\end{eqnarray}
where $G^{CC^{\prime}}_{\mu}(C,C^{\prime}=R,G,B=1,2,3)$ are gluons and $V^{15}_{\mu}$ is photon.
Then, the electric charge matrix of $V_{\mu}$ is
\begin{eqnarray}
Q_{V}=\left(\begin{array}{cccc}
0& 0 & 0 & -1/3\\
0& 0 & 0 & -1/3\\
0 & 0 & 0 & -1/3\\
1/3 & 1/3 & 1/3 & 0
\end{array}\right).
\end{eqnarray}
Three examples of the nonzero semi-leptonic Feynman diagrams in the tree level amplitudes are shown in Fig.~\ref{pic1}, where the Fig.~\ref{pic1}a and b are the t-channel and u-channel of
\begin{eqnarray}
\bar{u}_{C}+c_{C}\rightarrow e^{-}+\mu.
\end{eqnarray}
In addition, the Fig.~\ref{pic1}c is the s-channel of the quark lepton interaction
\begin{eqnarray}
c_{C}+\mu^{-}\rightarrow u_{C}+e^{-}.
\end{eqnarray}
The masses of $X^{\pm C}$ bosons must have been very large because the s, t and u-channels were still not observed.

\begin{figure}
\begin{center}
\includegraphics[width=72 mm]{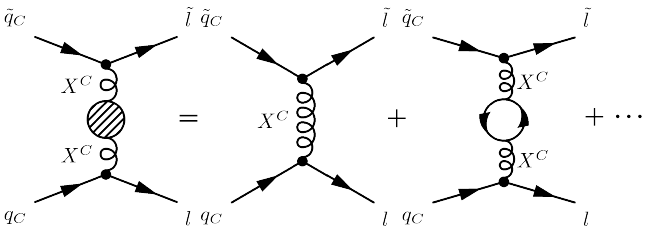}
\caption{\label{pic2} Amplitude of quark pair slips to lepton pair is zero because of electric charge conservation. The $q_{C}, \tilde{q}_{C}$ and $l,\tilde{l}$ are particular quarks and leptons. The vertexes in the diagram are described by Lagrangian (\ref{Lcolor}), especially Lagrangian \eqref{Xboson}.}
\end{center}
\end{figure}

The amplitudes in Fig.~\ref{pic2} are zero at least on the one-loop level in the model described by Lagrangian \eqref{Lcolor} because
\begin{eqnarray}\nonumber
M_{total} &\propto& \sum^{14}_{a=9}T^{a}_{LC}T^{a}_{LC}-\sum^{14}_{a,b=9}T^{a}_{LC}T^{a}_{CL}T^{b}_{LC}T^{b}_{LC}\\
&&-\sum^{14}_{a,b=9}T^{a}_{LC}T^{a}_{LC}T^{b}_{CL}T^{b}_{LC}+\cdots \\ \nonumber
&=&0-0-0+\cdots.
\end{eqnarray}
Note that all external fermions in Fig.\ref{pic2} are not anti-particles. The electric charge is not conserved in the process shown in Fig.~\ref{pic2} such that the total amplitude $M_{total}=0$. Which means electric charge conservation avoids quark pair slips to lepton pair in minimal coupling model (\ref{Lcolor}).

\section{Flavor mixing}
The left-handed flavor eigenstates of $d,s,b$ quark states can be defined as follows:
\begin{eqnarray}
-\frac{\sqrt{2}}{2}g \mathbf{Tr}[\bar{u}_{LC}\gamma^{\mu}d^{\prime}_{LC}W^{+}_{\mu}]|d^{\prime}_{LC}\rangle=\alpha_{1} |d^{\prime}_{LC}\rangle,\\
-\frac{\sqrt{2}}{2}g \mathbf{Tr}[\bar{c}_{LC}\gamma^{\mu}d^{\prime}_{LC}W^{+}_{\mu}]|s^{\prime}_{LC}\rangle=\alpha_{2} |s^{\prime}_{LC}\rangle,\\
-\frac{\sqrt{2}}{2}g \mathbf{Tr}[\bar{t}_{LC}\gamma^{\mu}d^{\prime}_{LC}W^{+}_{\mu}]|b^{\prime}_{LC}\rangle=\alpha_{3} |b^{\prime}_{LC}\rangle,
\end{eqnarray}
where $|d^{\prime}_{LC}\rangle, |s^{\prime}_{LC}\rangle$ and $|b^{\prime}_{LC}\rangle$ are flavor eigenstates of $d, s$ and $b$ quarks with left-handed chirality and $C$ color, respectively. 
The kinematic term of fermions in the Lagrangian (\ref{Lcolor}) is
\begin{eqnarray}
\mathbf{Tr}[i\bar{\Psi}\gamma^{\mu}\partial_{\mu}\Psi]=i\mathbf{Tr}[\bar{\Psi}_{L}\gamma^{\mu}\partial_{\mu}\Psi_{L}+\bar{\Psi}_{R}\gamma^{\mu}\partial_{\mu}\Psi_{R}].
\end{eqnarray}
The kinematic term of fermions can be decomposed as follows:
\begin{eqnarray}\nonumber
&&i\mathbf{Tr}[\bar{\Psi}_{L}\gamma^{\mu}\partial_{\mu}\Psi_{L}+\bar{\Psi}_{R}\gamma^{\mu}\partial_{\mu}\Psi_{R}]\\   \nonumber
&=&i\mathbf{Tr}\left[\sum_{C=R,G,B}\left[2\left(\bar{u}_{LC}\gamma^{\mu}\partial_{\mu} u_{LC}+ \bar{c}_{LC}\gamma^{\mu}\partial_{\mu} c_{LC} \right.\right.\right.\\  \nonumber
&&\left.\left.+\bar{t}_{LC}\gamma^{\mu}\partial_{\mu} t_{LC} \right)+\bar{d}^{\prime}_{LC}\gamma^{\mu}\partial_{\mu} d^{\prime}_{LC} \right]+\bar{e}_{L}\gamma^{\mu}\partial_{\mu} e_{L}\\   
&&\left.+\bar{\mu}_{L}\gamma^{\mu}\partial_{\mu} \mu_{L}+\bar{\tau}_{L}\gamma^{\mu}\partial_{\mu} \tau_{L}+\bar{\nu}^{\prime}_{L}\gamma^{\mu}\partial_{\mu} \nu^{\prime}_{L}+\{L \rightarrow R\}\right].\ \ \ \ \ 
\end{eqnarray}
The left-handed mass eigenstates of the $d, s$ and $b$ quarks are
\begin{eqnarray}
i\mathbf{Tr}\left[\bar{d}^{\prime}_{LC}\gamma^{\mu}\partial_{\mu} d^{\prime}_{LC}\right]|d_{LC}\rangle= m_{dL} |d_{LC}\rangle,\\
i\mathbf{Tr}\left[\bar{d}^{\prime}_{LC}\gamma^{\mu}\partial_{\mu} d^{\prime}_{LC}\right] |s_{LC}\rangle= m_{sL} |s_{LC}\rangle,\\
i\mathbf{Tr}\left[\bar{d}^{\prime}_{LC}\gamma^{\mu}\partial_{\mu} d^{\prime}_{LC}\right] |b_{LC}\rangle= m_{bL} |b_{LC}\rangle.
\end{eqnarray}
The CKM matrix is
\begin{eqnarray}
\left(\begin{array}{c}
|d^{\prime}_{LC}\rangle\\
|s^{\prime}_{LC}\rangle\\
|b^{\prime}_{LC}\rangle
\end{array}\right)
=\left(\begin{array}{ccc}
V_{ud}&V_{us}&V_{ub}\\
V_{cd}&V_{cs}&V_{cb}\\
V_{td}&V_{ts}&V_{tb}
\end{array}\right)
\left(\begin{array}{c}
|d_{LC}\rangle\\
|s_{LC}\rangle\\
|b_{LC}\rangle
\end{array}\right).
\end{eqnarray}
The right-handed $d, s$ and $b$ quark states can be defined after $L\rightarrow R$.

Similarly, the left-handed flavor eigenstates of neutrinos are
\begin{eqnarray}
-\frac{1}{2}g \mathbf{Tr}[\bar{e}_{L}\gamma^{\mu}\nu^{\prime}_{L}W^{+}_{\mu}]|\nu_{e L}\rangle=\alpha_{4} |\nu_{e L}\rangle,\\
-\frac{1}{2}g \mathbf{Tr}[\bar{\mu}_{L}\gamma^{\mu}\nu^{\prime}_{L}W^{+}_{\mu}]|\nu_{\mu L}\rangle=\alpha_{5} |\nu_{\mu L}\rangle,\\
-\frac{1}{2}g \mathbf{Tr}[\bar{\tau}_{L}\gamma^{\mu}\nu^{\prime}_{L}W^{+}_{\mu}]|\nu_{\tau L}\rangle=\alpha_{6} |\nu_{\tau L}\rangle.
\end{eqnarray}
The left-handed mass eigenstates of neutrinos are
\begin{eqnarray}
i\mathbf{Tr}\left[\bar{\nu}^{\prime}_{L}\gamma^{\mu}\partial_{\mu} \nu^{\prime}_{L}\right]|\nu_{1L}\rangle= m_{1L} |\nu_{1L}\rangle,\\
i\mathbf{Tr}\left[\bar{\nu}^{\prime}_{L}\gamma^{\mu}\partial_{\mu} \nu^{\prime}_{L}\right]|\nu_{2L}\rangle= m_{2L} |\nu_{2L}\rangle,\\
i\mathbf{Tr}\left[\bar{\nu}^{\prime}_{L}\gamma^{\mu}\partial_{\mu} \nu^{\prime}_{L}\right]|\nu_{3L}\rangle= m_{3L} |\nu_{3L}\rangle.
\end{eqnarray}
The PMNS matrix is
\begin{eqnarray}
\left(\begin{array}{c}
|\nu_{eL}\rangle\\
|\nu_{\mu L}\rangle\\
|\nu_{\tau L}\rangle
\end{array}\right)
=\left(\begin{array}{ccc}
U_{e1}&U_{e2}&U_{e3}\\
U_{\mu 1}&U_{\mu 2}&U_{\mu 3}\\
U_{\tau 1}&U_{\tau 2}&U_{\tau 3}
\end{array}\right)
\left(\begin{array}{c}
|\nu_{1L}\rangle\\
|\nu_{2L}\rangle\\
|\nu_{3L}\rangle
\end{array}\right).
\end{eqnarray}
The right-handed eigenstates of neutrinos can be defined similarly after $L\rightarrow R$.

\section{Effective total Lagrangian and gauge invariance}
\label{sec:1}
An effective total Lagrangian for color, flavor and Higgs interactions is
\begin{eqnarray}\nonumber
\label{Leffective}  
\mathcal{L}&=& \mathbf{Tr}\left[i\bar{\Psi}\gamma^{\mu}\partial_{\mu}\Psi+f \bar{\Psi} \gamma^{\mu}V_{\mu}\Psi-g\bar{\Psi}\gamma^{\mu} \Psi W_{\mu}\right.\\ \nonumber
&&+\bar{\Psi}\phi \Psi+V(\phi) -\frac{f^{2}}{2}{H}^{\mu\nu}H_{\mu\nu} -\frac{ g^{2} \xi}{2} {F}^{\mu\nu}F_{\mu\nu} \\ \nonumber 
&& -igF_{\mu\nu}\Psi^{\dagger}(\gamma^{\mu}\gamma^{\nu}-\gamma^{\nu\dagger}\gamma^{\mu\dagger})\Psi\\
&&\left.+i f \Psi^{\dagger}H_{\mu\nu}(\gamma^{\mu}\gamma^{\nu}-\gamma^{\nu\dagger}\gamma^{\mu\dagger})\Psi \right],
\end{eqnarray}
where $\phi$ is the Higgs field; $V(\phi)$ is Higgs potential; $f,g,\xi \in \mathbb{R}$ are coupling constants and the gauge field strength tensors are
\begin{eqnarray}
H_{\mu\nu}&=&\partial_{\mu}V_{\nu}-\partial_{\nu}V_{\mu}-ifV_{\mu}V_{\nu}+ifV_{\nu}V_{\mu},\\
F_{\mu\nu}&=&\partial_{\mu}W_{\nu}-\partial_{\nu}W_{\mu}-igW_{\mu}W_{\nu}+igW_{\nu}W_{\mu}.
\end{eqnarray}
The second line of Lagrangian (\ref{Leffective}) represents Yang-Mills theory terms, and the third line is magnetic moment terms.
Lagrangian \eqref{Leffective} is invariant under local gauge transformations of color space and flavor space rotation $\tilde{U}$ and $U$, respectively,
\begin{eqnarray}
\Psi^{\prime}=\tilde{U}\Psi U,
\end{eqnarray}
where 
\begin{eqnarray}
\tilde{U}\in SU(4^{\prime}),\quad  U \in SU(4),
\end{eqnarray}
 such that
\begin{eqnarray}
\gamma^{\mu\prime}&=&\tilde{U}\gamma^{\mu}\tilde{U}^{\dagger} \Rightarrow \gamma^{0\prime}\gamma^{\mu\prime}=\tilde{U}\gamma^{0}\gamma^{\mu}\tilde{U}^{\dagger},\\
V^{\prime}_{\mu}&=&\tilde{U}V_{\mu}\tilde{U}^{\dagger}-(\partial_{ \mu} \tilde{U}) \tilde{U}^{\dagger},\\
W^{\prime}_{\mu}&=&U^{\dagger} (\partial_{\mu}U)-U^{\dagger}W_{\mu}U.
\end{eqnarray}

\section{Gauge anomaly}
The Lagrangian (\ref{Leffective}) is flat space-time version of Yang-Mills theory (Pati-Salam type) in curved space-time and Einstein-Cartan gravity\cite{Li2014PatiSalamMI, Li:2022vaw}. The curved version theory has deep motivation from point of views of logic and geometry, which derived from square root metric and self-parallel transportation principle and quantized by sheaf quantization and path integral quantization. The anomaly in quantum field theory always means a symmetry is preserved in classial theory but violated in quantum version. The golbal symmetry anomaly might be accessed by quantum field theory, but the locally gauge symmetry anomaly (gauge anomaly) is belived to be a consistence condition for a gauge theory. We have to check the anomaly free condition for Pati-Salam model with this representation of fermions.

\begin{figure}
\begin{center}
\includegraphics[width=86 mm]{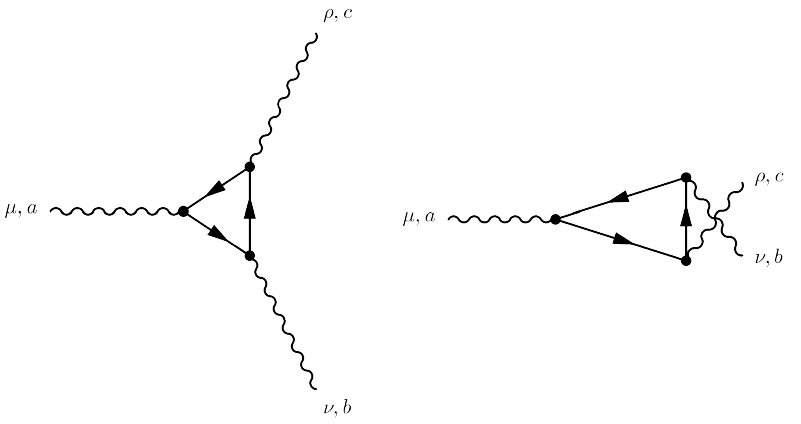}
\caption{\label{triangle} The Feynman diagrams about triangle anomaly. }
\end{center}
\end{figure}

In 4-dimenional space-time, the quantum gauge anomaly free condition can be checked by triangle Feymann diagram in Fig.~\ref{triangle}. 
The amplitude of Fig.~\ref{triangle} propotional to
\begin{eqnarray}\nonumber
iM^{abc\mu\nu\rho} \propto \mathbf{Tr}\left(T^{a}T^{b}T^{c}\right)+\mathbf{Tr}\left(T^{a}T^{c}T^{b}\right)\\
=2\mathbf{Tr}\left(T^{a}T^{(b}T^{c)}\right)=\frac{1}{2}d^{(abc)},
\end{eqnarray}
then for $SU(4)_{L}\times SU(4)_{R}$ chiral Yang-Mills theory, the current conservative equation has the formulation
\begin{eqnarray}
\partial_{\mu}J^{\mu,a}(x)\propto d^{(abc)}\epsilon^{[\mu\nu\rho\sigma]}F^{b}_{\mu\nu}F^{c}_{\rho\sigma},
\end{eqnarray}
such that the indices $bc$ satisfy commutation and anti-commutation relations
\begin{eqnarray}
d^{(a[bc])}=0.
\end{eqnarray}
The analyse about $SU(4^{\prime})$ color gauge Yang-Mills theory is similar. Note that a fermions loop cannot interact with flavor and color gauge bosons in one triangle anomaly Feymann diagram at the same time. So  the $SU(4)_{L}\times SU(4)_{R} \times SU(4^{\prime})$ Pati-Salam model is anomaly free.

\section{Monopole and the topology of space-time}
As an example, we choose $SU(4)_{L}\times SU(4)_{R}$ flavor gauge bosons to analyse the problem of monopole. We can combine the $SU(4)_{L}\times SU(4)_{R}$ minimal coupling, Yang-Mills and tolopogical terms of flavor gauge bosons as follow which related with monopole 
\begin{eqnarray}\label{monopole}
\mathcal{L}_{topology}=-g\bar{\Psi}\gamma^{\mu} \Psi W_{\mu}-\frac{g^{2} \xi}{2} {F}^{\mu\nu}F_{\mu\nu}-\frac{\eta g^{2} \xi}{2} \widetilde{F}^{\mu\nu}F_{\mu\nu},
\end{eqnarray}
where
\begin{eqnarray}
\widetilde{F}^{\mu\nu}=\epsilon^{\mu\nu\rho\sigma}F_{\rho\sigma}
\end{eqnarray}
are eletro-magnetic dual gauge strength tensor of $F_{\mu\nu}$. The Euler-Lagrangian equation of $W_{\mu}$ for the Lagrangian (\ref{monopole}) is
\begin{eqnarray}\label{total-em}
2g\xi\partial_{\mu}F^{\mu\nu}+2\eta g\xi\partial_{\mu}\widetilde{F}^{\mu\nu}=J^{\nu}, \\
 J^{\nu}=\bar{\Psi}\gamma^{\nu}\Psi=J^{\nu}_{e}+J^{\nu}_{m}.
\end{eqnarray}
We decompose the equation (\ref{total-em}) as follow
\begin{eqnarray}\label{eletrocurrent}
\partial_{\mu}F^{\mu\nu}=\frac{1}{2g\xi}J^{\nu}_{e}, \\  \label{monoplecurrent}
\partial_{\mu}\widetilde{F}^{\mu\nu}=\frac{1}{2\eta g\xi}J^{\nu}_{m},
\end{eqnarray}
where $J^{\nu}_{e}$ is electro current and $J^{\nu}_{m}$ is monopole current. The fundamental thing in quantum field theory is action $S$
\begin{eqnarray}
S=-\int_{M} \omega \left(\frac{\eta g^{2} \xi}{2} \widetilde{F}^{\mu\nu}F_{\mu\nu}\right),
\end{eqnarray}
where $\omega$ is volume form and $M$ is the base manifold of space-time.Note that the monopole related topological term in (\ref{monopole}) is the second Chern class, the action $S$ about the topological term only relies on the topological structure of the manifold $M$ and propotional with the second Chern number $C_{2}$
\begin{eqnarray}
S\propto C_{2}, &\quad& C_{2}\in \mathbb{Z}.
\end{eqnarray}
We can easily calulate the second Chern number $C_{2}$ with $M$ equals topologies $S^{4}$ and $S^{1}\times S^{3}$
\begin{eqnarray}
C_{2}=2 ,\quad M=S^{4},\\
C_{2}=0 , \quad M=S^{1}\times S^{3}.
\end{eqnarray}

\section{Conclusions and discussion}
Based on the gauge group $SU(4)_{L}\times SU(4)_{R} \times SU(4^{\prime})$ of Pati-Salam model, a representation of fermions is suggested in this paper. The boson-fermion-antifermion vertexes bring by the $SU(4)_{L}\times SU(4)_{R}$ chiral flavor and the $SU(4^{\prime})$ color gauge group were discussed. The electric charge of each particle was consistently defined, and a pair of possible CKM and PMNS matrix formulations were illustrated. An effective total Lagrangian of the model was given.

The experimental data restricts the masses of particles $X^{\pm C}, Y^{1},Y^{1}_{*},Y^{2}$ and $Y^{2}_{*}$ were superheavy. How the masses be generated for these particles requires further discussions.

\begin{acknowledgments}
We thank prof. Hong-Fei Zhang, Zhao Li, Cai-Dian Lv, Ming Zhong, Yong-Chang Huang and Chao-Guang Huang for valuable discussions. If without prof. Zhao Li kind helps, this work is impossible.
\end{acknowledgments}

\appendix

\section{Appendix A: Generators of $SU(4)$ group }
\label{AppendixA}
Generators of the $SU(4)$ group are as follows
\begin{eqnarray*}
\begin{array}{cc}
{T}^1=\frac{1}{2}\begin{pmatrix}
0&1&0&0\\
1&0&0&0\\
0&0&0&0\\
0&0&0&0\\
\end{pmatrix},&
{T}^2=\frac{1}{2}\begin{pmatrix}
0&-i&0&0\\
i&0&0&0\\
0&0&0&0\\
0&0&0&0\\
\end{pmatrix}, \\
{T}^3=\frac{1}{2}\begin{pmatrix}
1&0&0&0\\
0&-1&0&0\\
0&0&0&0\\
0&0&0&0\\
\end{pmatrix},&
{T}^4=\frac{1}{2}\begin{pmatrix}
0&0&1&0\\
0&0&0&0\\
1&0&0&0\\
0&0&0&0\\
\end{pmatrix},\\
{T}^5=\frac{1}{2}\begin{pmatrix}
0&0&-i&0\\
0&0&0&0\\
i&0&0&0\\
0&0&0&0\\
\end{pmatrix},&
{T}^6=\frac{1}{2}\begin{pmatrix}
0&0&0&0\\
0&0&1&0\\
0&1&0&0\\
0&0&0&0\\
\end{pmatrix}, \\
{T}^7=\frac{1}{2}\begin{pmatrix}
0&0&0&0\\
0&0&-i&0\\
0&i&0&0\\
0&0&0&0\\
\end{pmatrix},&
{T}^8=\frac{\sqrt{3}}{6}\begin{pmatrix}
1&0&0&0\\
0&1&0&0\\
0&0&-2&0\\
0&0&0&0\\
\end{pmatrix},  \\  
 {T}^9=\frac{1}{2}\begin{pmatrix}
0&0&0&1\\
0&0&0&0\\
0&0&0&0\\
1&0&0&0\\
\end{pmatrix}, & 
{T}^{10}=\frac{1}{2}\begin{pmatrix}
0&0&0&-i\\
0&0&0&0\\
0&0&0&0\\
i&0&0&0\\
\end{pmatrix},  \\  
{T}^{11}=\frac{1}{2}\begin{pmatrix}
0&0&0&0\\
0&0&0&1\\
0&0&0&0\\
0&1&0&0\\
\end{pmatrix},&
{T}^{12}=\frac{1}{2}\begin{pmatrix}
0&0&0&0\\
0&0&0&-i\\
0&0&0&0\\
0&i&0&0\\
\end{pmatrix},\\
{T}^{13}=\frac{1}{2}\begin{pmatrix}
0&0&0&0\\
0&0&0&0\\
0&0&0&1\\
0&0&1&0\\
\end{pmatrix},&
{T}^{14}=\frac{1}{2}\begin{pmatrix}
0&0&0&0\\
0&0&0&0\\
0&0&0&-i\\
0&0&i&0\\
\end{pmatrix}, \\   
 {T}^{15}=\frac{\sqrt{6}}{12}\begin{pmatrix}
1&0&0&0\\
0&1&0&0\\
0&0&1&0\\
0&0&0&-3\\
\end{pmatrix}.
\end{array}
\end{eqnarray*}

\section{Appendix B: The possibility of the chiral symmetry breaking of flavor gauge interaction}
\label{AppendixB}
$\gamma^{5}=i\gamma^{0}\gamma^{1}\gamma^{2}\gamma^{3}$ such that
\begin{eqnarray}
\gamma^{5\dagger}=\gamma^{5}.
\end{eqnarray}
The gamma matrices satisfy
\begin{eqnarray*}
\frac{1-\gamma^{5\dagger}}{2}\frac{1+\gamma^{5}}{2}=0,&& \frac{1+\gamma^{\dagger}}{2}\frac{1-\gamma^{5}}{2}=0,\\
\frac{1-\gamma^{5\dagger}}{2}\gamma^{0}\gamma^{\mu}\frac{1+\gamma^{5}}{2}=0,&& \frac{1+\gamma^{5\dagger}}{2}\gamma^{0}\gamma^{\mu}\frac{1-\gamma^{5}}{2}=0,
\end{eqnarray*}
such that Lagrangian $-g\bar{\Psi}\gamma^{\mu} \Psi W_{\mu}$ can be decomposed into two chiral components in any Dirac matrix representation, i.e.,
\begin{eqnarray*}\nonumber
-g\mathbf{Tr}\left[\bar{\Psi}\gamma^{\mu} \Psi W_{\mu}\right]=-g\mathbf{Tr}\left[\bar{\Psi}_{L}\gamma^{\mu} \Psi_{L} W_{\mu}^{a}{T}^{a}+\bar{\Psi}_{R}\gamma^{\mu} \Psi_{R} W_{\mu}^{a}{T}^{a}\right].
\end{eqnarray*}

\section{Appendix C: Cross sections}
\label{AppendixC}
\begin{figure}
\begin{center}
\includegraphics[width=53.8 mm]{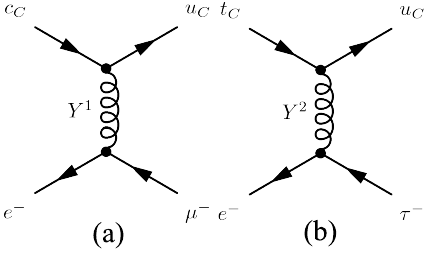}
\caption{\label{pic3}Examples of nonzero tree level amplitudes of the beyond SM FCNCs transported by neutral gauge bosons $Y^{1}$ and $Y^{2}$.}
\end{center}
\end{figure}

The cross section in Fig.~\ref{pic3}a can be represented as follows:
\begin{eqnarray}
|M|^{2}&=&\frac{16 g^{2}}{(t-m_{Y^{1}}^{2})^{2}}\left( (s-m_{e}^{2}-m_{\mu}^{2})(s-m_{u}^{2}-m_{c}^{2}) \right. \\ \nonumber
&&+(u-m_{e}^{2}-m_{\mu}^{2})(u-m_{u}^{2}-m_{c}^{2}) +8 m_{e}m_{\mu} m_{u}m_{c} \\ \nonumber
&& \left. +2m_{e}m_{\mu}(t -m_{u}^{2}-m_{c}^{2})+2m_{u}m_{c}(t -m_{e}^{2}-m_{\mu}^{2}) \right).
\end{eqnarray}
The cross section in Fig.~\ref{pic3}b replaces $m_{Y^{1}},m_{\mu}$ and $m_{c}$ with $m_{Y^{2}},m_{\tau}$ and $m_{t}$, respectively.

\begin{figure}
\begin{center}
\includegraphics[width=80 mm]{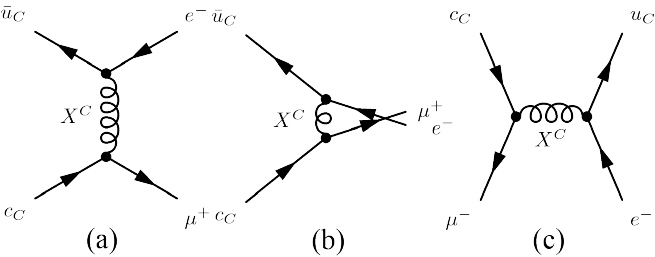}
\caption{\label{pic1} Nonzero tree-level amplitudes of semi-leptonic vertexes. The $e$ and $\mu$ are electron and mu leptons, the $u_{C}$ and $c_{C}$ are $u$ and $c$ quarks with color $C$. These processes were still not being observed in the experiments, indicating that the masses of $X$ bosons must have been superheavy. The $u,e$ and $c,\mu$ can be replaced with $t,\tau$ or $d^{\prime},\nu^{\prime}$ according to Lagrangian \eqref{Xboson}.}
\end{center}
\end{figure}

The cross sections in Fig.~\ref{pic1}a and b are
\begin{eqnarray}
|M_{t}|^{2}&=&\frac{288f^{4}}{(t-m_{X}^{2})^{2}}\left( (s-m_{\mu}^{2}-m_{c}^{2})(s-m_{e}^{2}-m_{u}^{2})\right.\\ \nonumber
&&+(u-m_{\mu}^{2}-m_{c}^{2})(u-m_{e}^{2}-m_{u}^{2})+8m_{e}m_{\mu}m_{u}m_{c}\\ \nonumber
&&\left. +2 m_{e}m_{\mu}(t-m_{\mu}^{2}-m_{c}^{2})+2 m_{\mu}m_{c}(t-m_{e}^{2}-m_{u}^{2})\right),\\
|M_{u}|^{2}&=&\frac{288f^{4}}{(u-m_{X}^{2})^{2}}\left( (s-m_{\mu}^{2}-m_{c}^{2})(s-m_{e}^{2}-m_{u}^{2})\right.\\ \nonumber
&&+(t-m_{\mu}^{2}-m_{c}^{2})(t-m_{e}^{2}-m_{u}^{2})+8m_{e}m_{\mu}m_{u}m_{c}\\ \nonumber
&&\left. +2 m_{e}m_{\mu}(t-m_{\mu}^{2}-m_{c}^{2})+2 m_{\mu}m_{c}(t-m_{e}^{2}-m_{u}^{2})\right),
\end{eqnarray}
where $m_{X}, m_{e},m_{\mu},m_{u}$ and $m_{c}$ are the masses of the $X^{\pm C}$ bosons, $e,\mu$ leptons, $u$ and $c$ quarks. The $s,t$ and $u$ are defined as follows:
\begin{eqnarray}
s&=&(p_{1}+p_{2})^{2}=(p_{3}+p_{4})^{2},\\
t&=&(p_{1}-p_{3})^{2}=(p_{2}-p_{4})^{2},\\
u&=&(p_{1}-p_{4})^{2}=(p_{2}-p_{3})^{2}.
\end{eqnarray}
The cross section of Fig.~\ref{pic1}c is given as follows:
\begin{eqnarray}
|M_{s}|^{2}&=&\frac{288f^{4}}{(s-m_{X}^{2})^{2}}\left( (t-m_{\mu}^{2}-m_{c}^{2})(t-m_{e}^{2}-m_{u}^{2})\right.\\ \nonumber
&&+(u-m_{\mu}^{2}-m_{c}^{2})(u-m_{e}^{2}-m_{u}^{2})+8m_{e}m_{\mu}m_{u}m_{c}\\ \nonumber
&&\left. +2 m_{e}m_{\mu}(s-m_{\mu}^{2}-m_{c}^{2})+2 m_{\mu}m_{c}(s-m_{e}^{2}-m_{u}^{2})\right)
.
\end{eqnarray}

\bibliographystyle{apsrev4-1}


\renewcommand{\baselinestretch}{1}
\normalsize

%

\end{document}